\documentclass[12pt]{article}

\usepackage[tbtags]{amsmath}
\usepackage{graphicx}
\usepackage{url}
\usepackage{amssymb}
\usepackage{cite}
\usepackage{multirow}
\usepackage{epsfig}
\usepackage{url}
\usepackage{subfig}
\begin{document}
\topmargin 0pt
\oddsidemargin 0mm
\renewcommand{\thefootnote}{\fnsymbol{footnote}}
\begin{titlepage}

\vspace{5mm}

\begin{center}
{\Large \bf Graphene and Some of its Structural
Analogues: Full-potential Density Functional
Theory Calculations} \\

\vspace{6mm} {\large Gautam Mukhopadhyay\footnote{Corresponding author's E-mail: gmukh@phy.iitb.ac.in; g.mukhopa@gmail.com} and Harihar Behera\footnote{E-mail: harihar@phy.iitb.ac.in;
behera.hh@gmail.com }}    \\
\vspace{5mm}
{\em Department of Physics, Indian Institute of Technology, Powai, Mumbai-400076, India} \\

\end{center}

\vspace{5mm}
\centerline{\bf {Abstract}}
\vspace{5mm}
  Using full-potential density functional calculations we have investigated the structural and electronic properties of graphene and some of its structural analogues, viz., monolayer (ML) of SiC, GeC, BN, AlN, GaN, ZnO, ZnS and ZnSe. While our calculations corroborate some of the reported results based on different methods, our results on ZnSe, the two dimensional bulk modulus of ML-GeC, ML-AlN, ML-GaN, ML-ZnO and ML-ZnS and the effective masses of the charge carriers in these binary mono-layers are something new. With the current progress in synthesis techniques, some of these new materials may be synthesized in near future for
applications in nano-devices.
 \\


\end{titlepage}

\section{Introduction}
Graphene is a crystal of carbon (C) atoms tightly
bound in a two dimensional (2D) hexagonal lattice.
It is a monolayer of carbon atoms (ML-C), i.e., one atom
thick. The exotic properties of this 2D material
were revealed only in 2004-2005 by a series of
papers coming from the Manchester \cite{1,2,3} and Columbia \cite{4} groups.
The unambiguous synthesis (by mechanical
exfoliation of graphite), identification (by
transmission electron microscopy (TEM)) and
experimental determination of some of the exotic
properties of graphene were reported first in 2004, by the
Manchester group led by Novoselov and Geim \cite{1}.
 In 2010, Konstantin S. Novoselov and Andre Geim were awarded the
Nobel Prize in Physics for the ``groundbreaking
experiments regarding the two-dimensional material
graphene". However, graphene research has a history
which dates back to the 1859 work of Brodie \cite{5, 6, 7, 8}.
The term "graphene" was introduced by
Boehm and his colleagues in 1986 \cite{9}. Graphene $=$ ``graph" $+$ ``ene"
and the term ``graph" is derived from the word ``graphite" and
the suffix ``ene" refers to polycyclic aromatic hydrocarbons. Now graphene is considered as the basic building block of graphitic materials (i.e.,
graphite $=$ stacked graphene, fullerenes $=$ wrapped graphene, nanotube $=$ rolled graphene, graphene nanoribbon $=$ nano-scale finite area sized
rectangular graphene). \\
\indent
The existence of free-standing 2D crystals were
believed impossible for several years, because they
would ultimately turn into a three-dimensional (3D)
objects as predicted by Peierls \cite{10},
Landau \cite{11} and Mermin \cite{12}. The theoretical reason for this is that at finite
temperatures, the displacements due to thermal
fluctuations could be of the same magnitude as the
inter-atomic distances, which make the crystal
unstable. Further, experimentally one generally
finds that thin films cannot be synthesized below a
certain thickness, due to islands formations or even
decomposition. Hence, the synthesis of graphene \cite{1}
was surprising which put a
question mark on the predictions of Peierls, Landau
and Mermin. However, this issue was (at least
partially) solved when it was shown that freestanding
graphene sheets display spontaneous
ripples owing to thermal fluctuation \cite{13},
and therefore real graphene is not perfectly flat. It is
important to note that such instabilities are the result
of thermal fluctuations which disappear at
temperature T $= 0^\circ $ K. This aspect will be used later
in our discussions on the study of the stability of
graphene and some of its structural analogues
reported here on the basis of their ground state (i.e.,
T $= 0^\circ $ K) total energies. \\
\indent
Being a one-atom-thick planar crystal of C atoms,
graphene is the thinnest nano-material ever known.
It has exotic mechanical, thermal, electronic, optical
and chemical properties, such as the high carrier
mobility, a weak dependence of mobility on carrier
concentration and temperature, unusual quantum
hall effect, hardness exceeding 100 times that of the
strongest steel of same thickness and yet flexible
(graphene can sustain elastic tensile strain more
than $20\%$ without breaking, and is brittle at certain
strain limit), high thermal conductivity comparable
to that of diamond and 10 times greater than that of
copper, negative coefficient of thermal expansion
over a wide range of temperature. Because of these
properties graphene has potentials for many novel
applications \cite{6, 7, 14, 15, 16, 17, 18, 19, 20 }.
The rapid advancements of nanotechnology and the computing power have
enabled the researchers in the field to explore the
unusual properties of graphene from many
perspectives of application and fundamental
science. The study of graphene is possibly the
largest and fastest growing field of research in
material science. \\
\indent
The impressive growth in the
research on graphene has inspired the study of other
graphene-like 2D materials \cite{20, 21}. For instance, a
number of 2D/quasi-2D nanocrystals (not based on
carbon) have been synthesized or predicted
theoretically in recent years. Representative samples
of other 2D nanocrystals which have been
synthesized include BN, MoS$_2$, MoSe$_2$, Bi$_2$Te$_3$ \cite{20, 21},
Si \cite{22}, ZnO \cite{23}. Recently, the density functional theory (DFT)
calculations of Freeman group \cite{24} have shown that when
the layer number of (0001)-oriented wurtzite (WZ) materials (e.g., AlN,
BeO, GaN, SiC, ZnO and ZnS) is small, the WZ
structures transform into a new form of stable
hexagonal BN-like structure. This prediction has
recently been confirmed in respect of ZnO \cite{23},
whose stability is attributed to the strong inplane
sp$^2$ hybridized bonds between Zn and O
atoms. Graphene-like 2D/quasi-2D honeycomb
structures of group-$IV$ and $III-V$ binary compounds
have also been studied \cite{25, 26} by using
pseudo-potential DFT calculations. Here, we report our
calculations on the structural and electronic properties
of graphene and some other graphene-like monolayer
(ML) structures of the binary compounds, viz.,
SiC, GeC, BN, AlN, GaN, ZnO, ZnS, ZnSe using the DFT.

\section{Computational Methods}
\begin{figure}[ht]
 \centering
\includegraphics[scale=0.65]{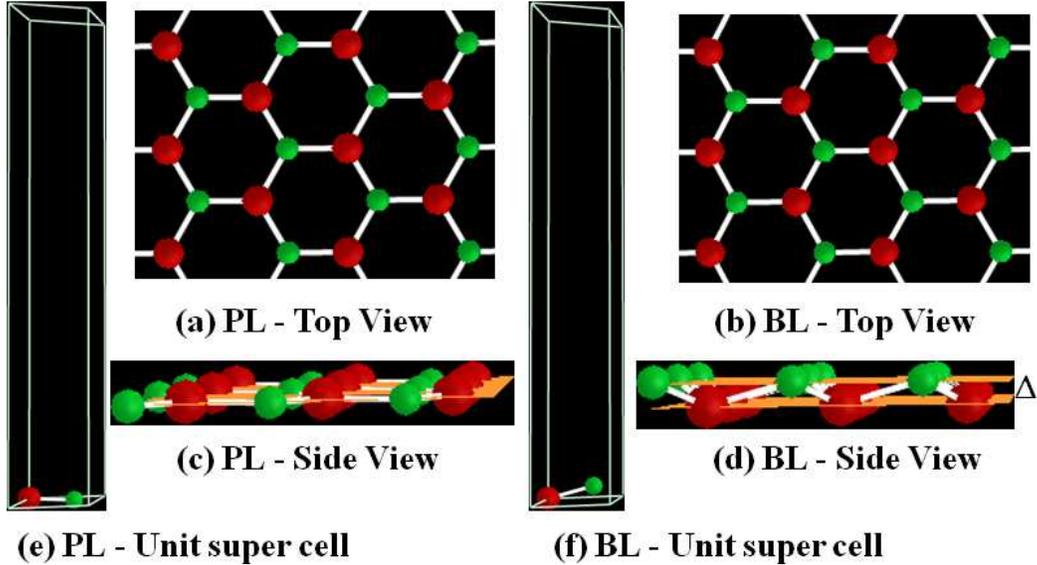}
\caption{Two possible structures of ML-BN: Planar (PL) and Buckled (BL) in ball-and-stick model. Sub-figures (a) and (b) show the top-down views of ML-BN in PL and BL structure respectively. In the buckled ML-BN, B atoms (large balls, red) and N atoms (small balls, green) are in two different parallel planes (sub-figure (d)), buckling parameter $\Delta$ is the perpendicular distance
between those parallel planes and $\Delta = 0$ \AA\, for a flat ML-BN (sub-figure (c)). Sub-figures (e) and (f) respectively depict the 3D
super cell of hexagonal BN in PL and BL structure with a large value of $c$-parameter (the length along the vertical axis).}
\end{figure}
We use the DFT based full-potential (linearized)
augmented plane wave plus local orbital 
(FP-(L)APW+lo) method \cite{27, 28, 29}
as implemented in the elk-code (http://
elk.sourceforge.net/) and the Perdew-Zunger
\cite{30} variant of local density
approximation (LDA) for our calculations. The
accuracy of this method and code has been
successfully tested in our previous studies
\cite{31, 32, 33, 34, 35, 36, 37}.
For plane wave expansion in the interstitial region,
we have used $8\leq |{\bf G} + {\bf k}|_{max}\times R_{mt} \leq 9$,
 where $R_{mt}$ is the smallest muffin-tin radius, for deciding the
plane wave cut-off. The Monkhorst-Pack \cite{38}
 $k$-point grid size of $20\times 20 \times 1$ was used for
 structural and of $30\times30\times 1$
for band structure and total density of states
(TDOS) calculations. The total energy was
converged within 2 $\mu$eV/atom. The 2D-hexagonal
structure was simulated by a 3D-hexagonal super
cell with a large value of $c$-parameter ($=|{\bf c}|= 40$ a.u.) as shown
in Figure 1. For structure optimization, we have considered two
different structures, viz., (i) Planar structure (PL)
and (ii) Buckled structure (BL) as shown in Figure
1 for a monolayer of BN (ML-BN), which is the
prototype of all the materials considered here (for
graphene, B and N atoms are to be replaced by the
C atoms).


\section{Results and Discussions}
We have optimized the 2D hexagonal structures
of graphene (ML-C) and the mono-layers of SiC
(ML-SiC), GeC (ML-GeC), BN (ML-BN), AlN
(ML-AlN), GaN (ML-GaN), ZnS (ML-ZnS) using
the principle of minimum energy for the stable
structure as per the following procedure. Initially we
assumed the planar structures for all the materials considered here.
 With this assumption, we calculated the ground state in-plane lattice
parameter $(a_0)$ of each of these structures as listed in
Table-1 along with the available reported values by
other authors. Then we investigated the possibility
of buckling in these structures at their assumed
planar ground states (characterized by their
respective $a_0$ values in planar states) by introducing
the concept of bucking (Figure 1) and the principle
of minimum energy for the most stable structure.
Our calculated variation of total energy (E) with
buckling parameter $(\Delta)$ at fixed value of $a_0$ was then
plotted to look for the value of $\Delta$ which corresponds
to the minimum energy.\\
\begin{figure}[ht]
 \centering
\includegraphics[scale=1.0]{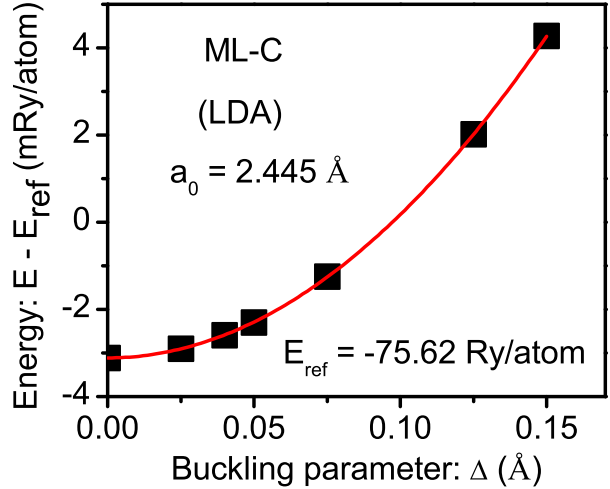}
\caption{Buckling probe of monolayer graphene (ML-C).}
\end{figure}
\begin{table}
\caption{Calculated LDA values of ground state in-plane lattice constant $a_0 (= |{\bf a}| = |{\bf b}|)$, buckling parameter $\Delta$ of monolayer graphene (MLC)
and some structurally similar binary compounds compared with available calculated data. PAW stands for the projector augmented wave.}
\centering
\begin{tabular}{|p{2.5cm}|p{1.5cm}|p{2.5cm}|p{6.5cm}|}
\hline
 Material (2D)& $a_0$ (\AA) & $\Delta$ (\AA) &   Remark/Reference \\
\hline
{}            &  {2.445}  &  0.0  &  Our Work/\cite{31,33}   \\
{ML-C}        &  {2.46}   &  0.0  &  PAW-potential/\cite{25}  \\
{}            &  {2.4431} &  0.0  &  Pseudo-potential/\cite{26}  \\
\hline
{}            &  {3.066}  &  0.0  &  Our Work/\cite{37}   \\
{ML-SiC}      &  {3.07}   &  0.0  &  PAW-potential/\cite{25}  \\
{}            &  {3.0531} &  0.0  &  Pseudo-potential/\cite{26}  \\
\hline
{ML-GeC}      &  {3.195}  &  0.0  &  Our Work/\cite{37}   \\
{}            &  {3.22}   &  0.0  &  PAW-potential/\cite{25}  \\
\hline
{}            &  {2.488}  &  0.0  &  Our Work/\cite{33, 34}   \\
{ML-BN}       &  {2.51}   &  0.0  &  PAW-potential/\cite{25}  \\
{}            &  {2.4870} &  0.0  &  Pseudo-potential/\cite{26}  \\
\hline
{ML-AlN}      &  {3.09}   &  0.0  &  Our Work/\cite{34}   \\
{}            &  {3.09}   &  0.0  &  PAW-potential/\cite{25}  \\
\hline
{ML-GaN}      &  {3.156}  &  0.0  &  Our Work/\cite{37}   \\
{}            &  {3.20}   &  0.0  &  PAW-potential/\cite{25}  \\
\hline
{ML-ZnO}      &  {3.20}   &  0.0 (used) &  Our Work/\cite{35}   \\
{}            &  {3.283}  &  0.0  &  PAW-potential/\cite{46}  \\
\hline
{ML-ZnS}      &  {3.7995} &  0.0  &  Our Work/\cite{36}   \\
{}            &  {3.890}  &  0.0  &  PAW-potential with GGA\cite{39}/\cite{47}  \\
\hline
{ML-ZnSe}     &  {3.996} &  0.0 (used) &  Our Work/\cite{37}   \\
\hline
\end{tabular}
\end{table}
\begin{figure}[]
 \centering
\includegraphics[scale=1.0]{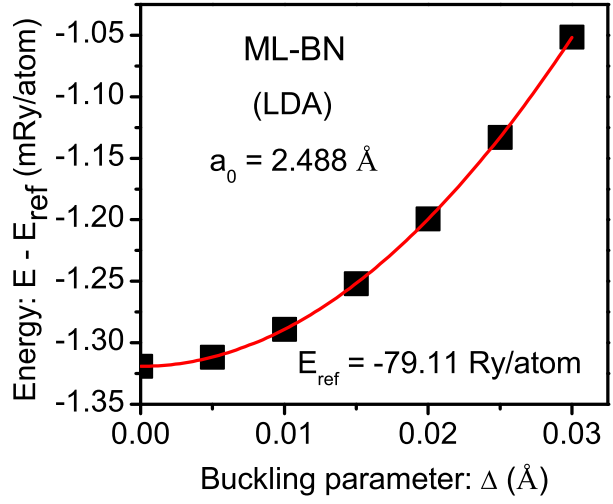}
\caption{Buckling probe of monolayer BN (ML-BN).}
\end{figure}
\begin{figure}[h]
 \centering
\includegraphics[scale=1.1]{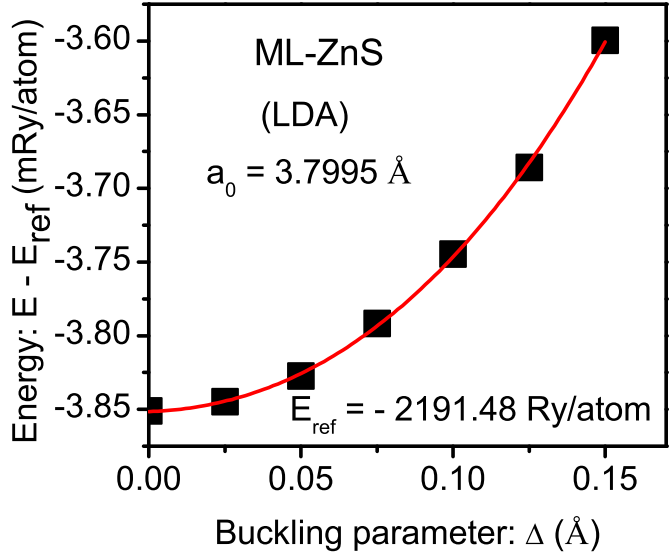}
\caption{Buckling probe of monolayer ZnS (ML-ZnS).}
\end{figure}
\begin{figure}[ht]
 \centering
\includegraphics[scale=1.3]{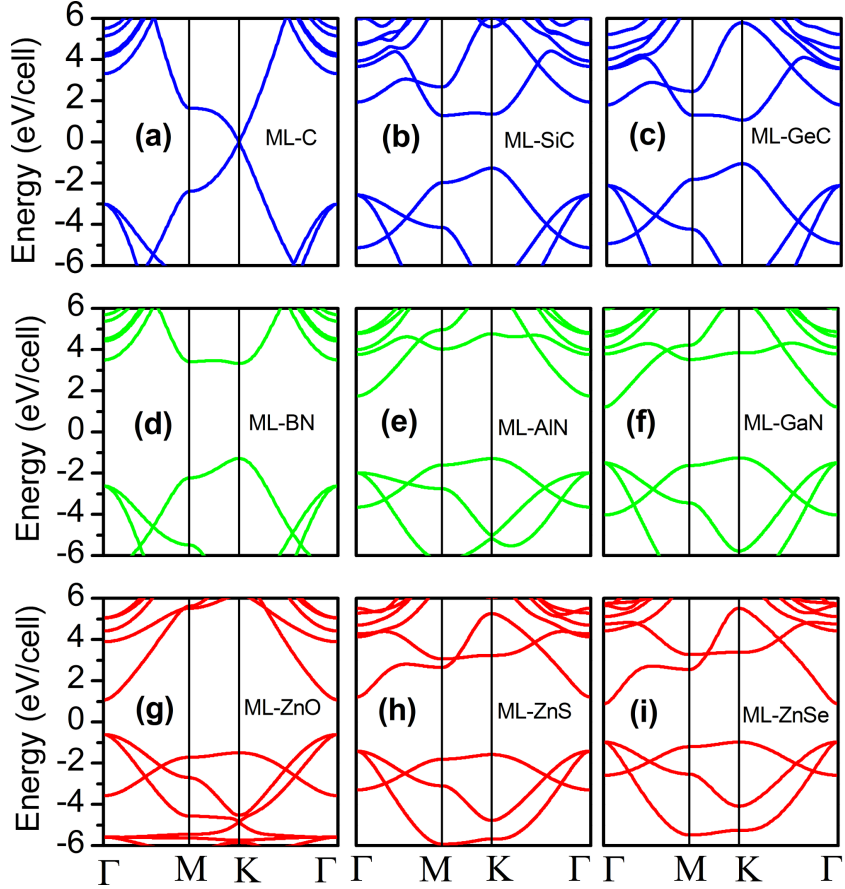}
\caption{Band structures of graphene (ML-C) and some other mono-layers of SiC, GeC, BN, AlN, GaN, ZnO, ZnS and ZnSe in graphene-like planar structure within LDA.}
\end{figure}

\begin{figure}
 \centering
\includegraphics[scale=1.5]{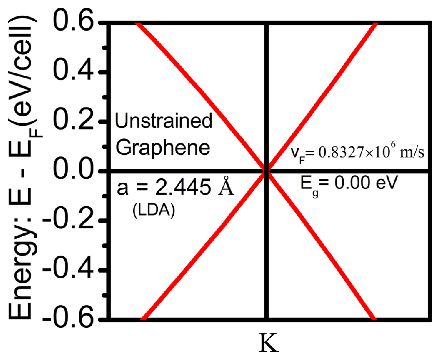}
\caption{Linear energy dispersion near the K point in monolayer
graphene's band structure.}
\end{figure}

\begin{table}
\caption{Calculated 2D bulk modulus $(B_{2D})$ value of monolayer graphene and some other mono-layers of binary compounds in graphene-like structure compared with available calculated data using other methods. Our cited values of $B_{2D}$ from Wang \cite{26} were obtained from Wang's calculated values of 2D elastic constants $\gamma _{11}$ and $\gamma _{12}$ and another definition of
$B_{2D} = (\gamma _{11} + \gamma _{12})/2$ \cite{49}.}
\centering
\begin{tabular}{|p{2.5cm}|p{2.5cm}|p{6.5cm}|}
\hline
 Material (2D)& $B_{2D}$ (N/m) &   Remark/Reference \\
\hline
{ML-C}        &  {223.85}    &  This Work   \\
{}            &  {214.41}    &  Pseudo-potential/\cite{26}  \\
{}            &  {220.00}    &  Pseudo-potential/\cite{48}  \\
\hline
{ML-SiC}      &  {125.66}    &  This Work    \\
{}            &  {121.945}   &  Pseudo-potential/\cite{26} \\
\hline
{ML-GeC}      &  {113.39}    &  This Work   \\
\hline
{ML-BN}       &  {188.03}    &  This Work   \\
{}            &  {181.91}    &  Pseudo-potential/\cite{26}  \\
{}            &  {212.0}     &  Born's Perturbation Method/\cite{49}  \\
\hline
{ML-AlN}      &  {113.44}      &  This Work  \\
\hline
{ML-GaN}      &  {109.45}     &  This Work   \\
\hline
{ML-ZnO}      &  {94.78}      & This Work   \\
\hline
{ML-ZnS}      &  {53.94}     &  This Work   \\
\hline
{ML-ZnSe}     &  {48.57}     &  This Work   \\
\hline
\end{tabular}
\end{table}

In Figures 2-4, we present our calculated results
selecting one example from each group of materials
considered here: graphene (ML-C) for Group-IV
material, ML-BN for Group-III-V material and MLZnS
for Group-II-VI material. As seen in the
Figures 2-4, these structures have minimum energy
at $\Delta = 0.00$ \AA \,, which means that these materials
adopt 2D planar structures in the ground state
(T $= 0^\circ$ K) unlike the case with silicene (graphene
analogue of Si) \cite{22, 31} that adopts a buckled
structure in its ground state. It is to be noted that
these results do not conflict with the theory of
stability of 2D crystals \cite{10, 11, 12}. Our calculated optimized
structural parameters of the materials are listed in
Table 1 along with available reported results based on other methods.
The stability tests ML-ZnO and ML-ZnSe are not yet complete.\\
\indent
Considering the facts that LDA usually underestimates
and GGA \cite{39} usually over-estimates the lattice constant,
the calculated value of $a_0$ slightly depends on the $c$-parameter used
for super cells in DFT based calculations as we have
seen before \cite{31} for graphene ($0.12\%$ lower value of $a$ for $c \rightarrow \infty)$ and silicene ($0.075\%$ lower value of $a$ for $c \rightarrow \infty$), and
the different methods of study used by different
authors, our results in Table 1 are acceptable.\\
\indent
Using the definition of the 2D bulk modulus of
planar a graphene-like 2D material as
$B_{2D} = A(\partial ^2E/\partial A^2)|_{A_{min}}$ (where A is the area of the
periodic cell of the 2D lattice and $A_{min}$ is the area
with minimum energy), we have calculated the
bulk modulus of each of the materials considered
here. These calculated values of $B_{2D}$ are listed in
Table 2 along with the available calculated data
based on other methods. As seen in Table 2, the
2D bulk modulus of graphene has the highest
value and that of ML-ZnSe has the lowest value
among the materials considered here, implying the
fact that graphene has stronger in-plane bonds
than the others and ML-ZnSe has the weakest
bond among these materials. ML-GeC is seen to
have same strength as that of ML-AlN. These data
may be useful in selecting the 2D materials having
desired level of strength in nano-mechanical
applications in specific situations.\\
\begin{table}
\caption{Calculated LDA band gaps compared and listed with reported values. }
\centering
\begin{tabular}{|p{2.5cm}|p{3.5cm}|p{6.5cm}|}
\hline
 Material (2D)& $E_g$ (eV) &    Remark/Reference \\
\hline
{ML-C}        &  {0} $(K\rightarrow K)$    &  Our Work/\cite{31,33}   \\
{}            &  {0} $(K\rightarrow K)$    &  PAW-potential/\cite{25}  \\
{}            &  {0}                       &  Pseudo-potential/\cite{26}  \\
\hline
{ML-SiC}      &  {2.547}$(K\rightarrow M)$  &  Our Work/\cite{37}   \\
{}            &  {2.52} $(K\rightarrow M)$  &  PAW-potential/\cite{25}  \\
{}            &  {4.19} $(K\rightarrow M)$  &  PAW-potential (LDA + GW$_0$)/\cite{25}\\
{}            &  {2.39}                     &  Pseudo-potential/\cite{26}  \\
\hline
{ML-GeC}      &  {2.108}$(K\rightarrow K)$  &  Our Work/\cite{37}   \\
{}            &  {2.09} $(K\rightarrow K)$  &  PAW-potential/\cite{25}  \\
{}            &  {3.83} $(K\rightarrow K)$  &  PAW-potential (LDA + GW$_0$)/\cite{25}\\
\hline
{ML-BN}       &  {4.606}$(K\rightarrow K)$  &  Our Work/\cite{33, 34}   \\
{}            &  {4.61} $(K\rightarrow K)$  &  PAW-potential/\cite{25}  \\
{}            &  {6.86} $(K\rightarrow \Gamma)$  &  PAW-potential (LDA + GW$_0$)/\cite{25}\\
{}            &  {4.35}                     &  Pseudo-potential/\cite{26}  \\
{}            &  {5.971}  (direct)           & Experiment/\cite{40}  \\
\hline
{ML-AlN}      &  {3.037}$(K\rightarrow \Gamma)$ & Our Work/\cite{34}      \\
{}            &  {3.08} $(K\rightarrow \Gamma)$ & PAW-potential/\cite{25}  \\
{}            &  {5.57} $(K\rightarrow \Gamma)$ &  PAW-potential (LDA + GW$_0$)/\cite{25}\\
\hline
{ML-GaN}      &  {2.462}$(K\rightarrow \Gamma)$ &  Our Work/\cite{37}   \\
{}            &  {2.27}$(K\rightarrow \Gamma)$  &  PAW-potential/\cite{25}  \\
\hline
{ML-ZnO}      &  {1.680}$(\Gamma\rightarrow \Gamma)$ &  Our Work/\cite{35}   \\
{}            &  {1.68} $(\Gamma\rightarrow \Gamma)$ & PAW-potential/\cite{46}  \\
\hline
{ML-ZnS}      &  {2.622}$(\Gamma\rightarrow \Gamma)$ &  Our Work/\cite{36}   \\
{}            &  {2.07} $(\Gamma\rightarrow \Gamma)$ &  PAW-potential with GGA\cite{39}/\cite{47}  \\
\hline
{ML-ZnSe}     &  {1.866} $(\Gamma\rightarrow \Gamma)$&  Our Work/\cite{37}   \\
\hline
\end{tabular}
\end{table}
\begin{table}
\caption{Calculated values of effective masses of the charge carriers in
ML-C, ML-Si, ML-Ge, ML-BN, ML-AlN, ML-GaN, ZnO, ZnS, ZnSe. $m_0$ is the rest mass
of electron.}
\centering
\begin{tabular}{|p{2.5cm}|p{3.5cm}|p{6.5cm}|}
\hline
 Material (2D)& Effective mass $(m_0)$ &    Remark/Reference \\
\hline
{ML-C}        &  {$m_e(K)  = 0$}                &  This Work  \\
{}            &  {$m_h(K)  = 0$}                &  This Work \\
\hline
{ML-SiC}      &  {$m_e(M)  = 0.411$}            &  This Work    \\
{}            &  {$m_h(K)  = - 0.488$}          &  This Work  \\
\hline
{ML-GeC}      & {$m_e(K)  = 0.509$}            &  This Work    \\
{}            & {$m_h(K)  = - 0.400$}          & This Work  \\
\hline
{ML-BN}       &  {$m_e(K)  = 0.920$}          &  This Work   \\
{}            &  {$m_h(K)  = - 0.617$}        &  This Work  \\
\hline
{ML-AlN}      &  {$m_e(\Gamma)  = 0.523$}     & This Work      \\
{}            &  {$m_h(K)  = - 1.470$}        & This Work  \\
\hline
{ML-GaN}      &  {$m_e(\Gamma)  = 0.266$}      &  This Work   \\
{}            &  {$m_h(K)  = - 1.157$}         &  This Work  \\
\hline
{ML-ZnO}      &  {$m_e(\Gamma)      = 0.253$}  &  This Work   \\
{}            &  {$m_{lh}(\Gamma)  = - 0.374$}     & This Work  \\
{}            &  {$m_{hh}(\Gamma)  = - 0.793$}     & This Work  \\
\hline
{ML-ZnS}      &  {$m_e(\Gamma)      = 0.173$}  &  This Work   \\
{}            &  {$m_{lh}(\Gamma)  = - 0.154$} &  This Work   \\
{}            &  {$m_{hh}(\Gamma)  = - 0.665$}     & This Work  \\
\hline
{ML-ZnSe}     &  {$m_e(\Gamma)      = 0.107$}  &  This Work  \\
{}            &  {$m_{lh}(\Gamma)  = - 0.103$} &  This Work   \\
{}            &  {$m_{hh}(\Gamma)  = - 0.652$}     & This Work  \\
\hline
\end{tabular}
\end{table}
\indent
The LDA band structures of graphene (ML-C)
and some other mono-layers of SiC, GeC, BN,
AlN, GaN, ZnO, ZnS and ZnSe in graphene-like
planar structure are depicted in Figure 5 and in
Table 3, we have listed our calculated LDA values
of the band gaps of these materials along with the
reported values. As seen in the explicit case of
ML-BN in Table 3, although LDA under-estimates
the band gap (in our case the calculated value is
about $23\%$ less than the experimental value of
5.971 eV) it correctly predicts the nature of the
gap. However, the use of a more advanced
approximation, such as the LDA + GW$_0$ used in
\cite{25}, improves the band gap
problem (in the case of ML-BN although it
overestimates band gap by $15\%$ in this case), but
the authors of \cite{25} have presented
their result in terms of an indirect band gap for
ML-BN as opposed to the experimental
observation of direct band gap in ref. \cite{40}.
In all other cases, the LDA nature of band
gap is the same as that of the LDA + GW$_0$ band
gap. The important point is that the actual band
gap is always more than the LDA value. \\
\indent
In Table 4, we present our calculated values of the
effective masses of the charge carriers in ML-C,
ML-Si, ML-Ge, ML-BN, ML-AlN, ML-GaN, ZnO,
ZnS, ZnSe, determined at the band edges at the
special points as appropriate for the material under
study. From the linear energy dispersion close to the
$K$ point of the hexagonal Brillouin Zone (BZ) as
shown in Figure 6, mass-less carriers in graphene
were inferred. As seen in Figure 6, near the $K$ point of the BZ,
the energy bad dispersion is linear in $k$:
\begin{equation}
E_{\pm} = \pm v_F \hbar k
\end{equation}
where $v_F$ is the magnitude of Fermi velocity of the
charge carriers in graphene and $(\hbar k)$ is the
magnitude of the momentum. Thus, charge carriers
in graphene behave like mass-less relativistic
particles. From the slope of the linear bands one
obtains the value of $v_F$, which in our calculation
corresponds to $v_F = 0.8327\times 10^6$ m/s. This value of $v_F$
is close to the experimentally measured value of $v_F
= 0.79 \times 10^6$ m/s in graphene monolayer deposited
on graphite substrate \cite{41} and the
reported calculated LAPW value of $0.833\times 10^6$ m/s
\cite{42} and about $20\%$ less than that
 measured in coupled multi-layers \cite{43, 44} and the tight-binding
value: $v_F = 1\times 10^6$ m/s \cite{15}.
The smallness in the value of $v_F$ in ref. \cite{41} was attributed to the electron-phonon interactions due to strong coupling with the graphite
substrate. However, the closeness of our calculated
value of $v_F$ for a freestanding graphene monolayer
with the results of ref. \cite{41} suggests a
very weak coupling of graphene to the graphite
substrate used in ref. \cite{41}. Surprisingly,
our calculated value of $v_F$ is close to the measured
value of $v_F = 0.81 \times 10^6$ m/s \cite{45} in
metallic single-walled carbon nanotube.



\section{Conclusions}
 Using full potential DFT calculations we have
investigated the structural and electronic properties
of graphene and some other graphene-like materials.
While our results corroborate the previous
theoretical studies based on different methods, our
calculations on ML-ZnSe, the two-dimensional bulk
modulus of ML-GeC, ML-AlN, ML-GaN, MLZnO, ML-ZnS, and the effective
masses of the binary monolayer compounds considered here are
our new results. We hope, with the advancement of
fabrication techniques, the hypothetical graphenelike
materials discussed here will be synthesized in
the near future for potential applications in a variety
of novel nano-devices.



\begin{thebibliography}{00}
\bibitem{1} Novoselov K.S., Geim A.K., Morozov S.V., Jiang D., Zhang Y.,
Dubono S.V., Grigorieva, I.V. and Firsov A.A., 2004.
Electric field effect in atomically thin carbon films. \emph{Science}
{\bf 306}, 666-669.
\bibitem{2} Novoselov K.S., Geim A. K., Morozov S.V., Jiang D., Zhang
Y., Dubonos S. V. and Firsov A. A., 2005. Twodimensional
gas of massless dirac fermions in graphene.
\emph{Nature} {\bf 438}, 197-200.
\bibitem{3} Novoselov K.S., Jiang D., Schedin F., Booth T., Khotkevich
V. V., Morozov S. V., Geim A. K., 2005, Twodimensional atomic crystals. \emph{Proc. Natl. Acad. Sci. USA.} {\bf 102}, 51-53.
\bibitem{4} Zhang Y., Tan Y-W., Stormer H. L. and Kim P., 2005. Experimental observation of the quantum hall effect and berry's phase in graphene. \emph{Nature} {\bf 438}, 201-204.
\bibitem{5} Brodie B.C., 1859. On the atomic weight of graphite. \emph{Phil.
Trans. R. Soc. Lond.} {\bf 149}, 249-259.
\bibitem{6} Geim A. K., 2011. Nobel Lecture: Random walk to graphene.
\emph{Rev. Mod. Phys.} {\bf 83}, 851-862.
\bibitem{7} Novoselov K.S., 2011. Nobel lecture: Graphene: materials in
the flatland. \emph{Rev. Mod. Phys.} {\bf 83}, 837-849.
\bibitem{8} Geim A. K., 2012. Graphene prehistory. {\emph Phys. Scr.} {\bf T146},
014003-014007.
\bibitem{9} Boehm H.P., Setton R. and Stumpp E., 1986. Nomenclature and
terminology of graphite intercalation compounds. \emph{Carbon} {\bf 4},
241-245.
\bibitem{10} Peierls R., 1935. Quelques proprieties typiques descorpses
solides. \emph{Ann. Inst. Henri Poincare} {\bf 3}, 177-222.
\bibitem{11} Landau L.D., 1937. Zur theorie der phasenumwandlungen II.
\emph{Phys. Z. Sowjetunion} {\bf 11}, 26-35.
\bibitem{12} Mermin N.D.,1968, Crystalline order in two dmension. \emph{Phys.
Rev.} {\bf 176}, 250-254.
\bibitem{13} Meyer J.C., Geim A.K., Katsnelson M.I., Novoselov K.S.,
Booth T.J. and Roth S., 2007. The structure of suspended
graphene sheets. \emph{Nature} {\bf 446}, 60-63.
\bibitem{14} Geim A. K. and Novoselov K.S., 2007. The rise of graphene.
\emph{Nature Mater.} {\bf 6}, 183-191.
\bibitem{15} Castro Neto A.H., Guinea F., Peres N.M.R., Novoselov K.S.
and Geim A.K., 2009. The electronic properties of graphene.
\emph{Rev. Mod. Phys.} {\bf 81}, 109-162.
\bibitem{16} Geim A. K., 2009. Graphene: Status and prospects. \emph{Science} {\bf 324}, 1530-1534.
\bibitem{17} Abergel D.S.L., Apalkov V., Berashevich J., Ziegler K. and
Chakraborty T., 2010. Properties of graphene: a theoretical
perspective. \emph{Adv. in Phys.} {\bf 59}, 261-482.
\bibitem{18} Singh V., Joung D., Zhai L., Das S., Khondaker S. I. and Seal
S., 2011. Graphene based materials: Past, present and future.
\emph{Prog. in Mater. Sci.} {\bf 56}, 1178-1271.
\bibitem{19} Castro Neto A.H. and Novoselov K., 2011. New directions in
science and technology: two-dimensional crystals. \emph{Rep. Prog. Phys.}
{\bf 74}, 1-9.
\bibitem{20} Novoselov K.S., Jiang D., Schedin F., Booth T., Khotkevich
V. V., Morozov S. V., Geim A. K., 2005b, Twodimensional
atomic crystals. \emph{Proc. Natl. Acad. Sci. USA.} {\bf 102}, 51-53.
\bibitem{21} Castro Neto A.H. and Novoselov K., 2011. New directions in
science and technology: two-dimensional crystals. \emph{Rep.
Prog. Phys.} {\bf 74}, 1-9.
\bibitem{22} Kara A., Enriquez H., Seitsonen A.P., Lew Yan Voon L.C. and
Vizzini S., 2012. A review on silicene-new candidate for
electronics. \emph{Surf. Sci. Rep.} {\bf 67}, 1-18.
\bibitem{23} Tusche C., Meyerheim H. L. and Kirshner J., 2007. Observation
of depolarized ZnO(0001) monolayers: Formation of unreconstructed planar sheets. \emph{Phys. Rev. Lett.} {\bf 99}, 026102 (4 pp).
\bibitem{24} Freeman C.L., Claeyssens F., AllanN. L. and Harding J. H.,
2006. Graphitic nanofilms as precursors to wurtzite films: Theory.
\emph{Phys. Rev. Lett.} {\bf 96}, 066102 (4 pp).
\bibitem{25} \c{S}ahin H., Cahangirov S., Topsakal M., Bekaroglu E., Akt\"{u}rk E.,
Senger R.T. and Ciraci S., 2009. Monolayer honeycomb structures of group-IV and III-V binary compounds: First-principles calculations. \emph{Phys. Rev. B} {\bf 80}, 155453 (12 pp).
\bibitem{26} Wang S., 2010. Studies of physical properties of twodimensional
hexagonal crystals by first-principles calculations. \emph{J. Phys. Soc. Jpn}
{\bf 79(6)}, 064602-064607.
\bibitem{27} Sj\"{o}stedt E., Nordstr\"{o}m L. and Singh D. J., 2000. An alternative
way of linearizing the augmented plane- wave method. \emph{Solid State Commun.} {\bf 114}, 15-20.
\bibitem{28} Madsen G.K.H., Blaha P., Schwarz K., Sj\"{o}stedt E. and
Nordstr\"{o}m L., 2001. Efficient linearization of the augmented
plane-wave method. \emph{Phys. Rev. B} {\bf 64}, 195134 (9 pp).
\bibitem{29} Singh D. J. and Nordstr\"{o}m L., 2006. Planewaves,
pseudopotentials, and the LAPW method (Springer, N.Y.).
\bibitem{30} Perdew P. and Zunger A., 1981. Self-interation correction to
density-functional approximations for many electron
systems. \emph{Phys. Rev. B} {\bf 23}, 5048-5079.
\bibitem{31} Behera H. and Mukhopadhyay G., 2010. Structural and
electronic properties of graphene and silicene: An FP-(L)APW+lo Study. \emph{AIP Conf. Proc.} {\bf 1313}, 152-155. \url{http://arxiv.org/abs/1111.1282}
\bibitem{32} Behera H. and Mukhopadhyay G., 2011. First-principles study
of structural and electronic properties of germanene. \emph{AIP
Conf. Proc.} {\bf 1349}, 823-824. \url{http://arxiv.org/abs/1111.6333}
\bibitem{33} Behera H. and Mukhopadhyay G., 2012. Strain-tunable band
gap in graphene/h-BN hetero-bilayer. \emph{J. Phys.Chem. Solids} {\bf 73}, 818-821.
\url{http://arxiv.org/abs/1204.2030}
\bibitem{34} Behera H. and Mukhopadhyay G., 2012. Strain-tunable bandgaps
of two-dimensional hexagonal BN and AlN: An FP-(L)APW+lo study.
\emph{AIP Conf. Proc.} {\bf 1447}, 273-274. \url{http://arxiv.org/abs/1206.3162}
\bibitem{35} Behera H. and Mukhopadhyay G., 2012. Strain-tunable direct
band gap of ZnO monolayer in graphene-like honeycomb structure.
\emph{Phys. Lett. A.} {\bf 376}, 3287-3289. \url{http://arxiv.org/abs/1211.3034}
\bibitem{36} Behera H. and Mukhopadhyay G., 2012. Strain-tunable band
gap of a monolayer graphene analogue of ZnS monolayer. Presented in the $20^{th}$ International Conference on Composites/Nano-Engineering, Beijing, China. \url{http://arxiv.org/abs/1210.3309}
\bibitem{37} Mukhopadhyay G. and Behera H., 2012, Structural and
electronic properties of graphene and graphene-like
materials. Presented in the $20^{th}$ International Conference on Composites/Nano-Engineering, Beijing, China.
\bibitem{38} Monkhorst H.J. and Pack J.D., 1976. Special points for
Brillouin-zone integrations. \emph{Phys. Rev. B} {\bf 13}, 5188-5192.
\bibitem{39} Perdew J. P., Burke K. and Wang Y., 1996. Generalized
gradient approximation for the exchange-correlation hole of
a many-electron system. \emph{Phys. Rev. B} {\bf 54}, 16533-16539.
\bibitem{40} Watanabe K., Taniguchi T. and Kanda H., 2004. Direct bandgap
properties and evidence for ultraviolet lasing of hexagonal boron nitride single crystal. \emph{Nature Mater} {\bf 3}, 404-409.
\bibitem{41} Li G., Luican A. and Andrei E.Y., 2009. Scanning tunneling
spectroscopy of graphene on graphite. \emph{Phys. Rev. Lett.} {\bf 102},
176804 (4 pp).
\bibitem{42} Gmitra M., Konschuh S., Ertler C., Ambrosch-Draxl C. and
Fabian J., 2009. Band-structure topologies of graphene: Spin-orbit coupling effects from first principles, \emph{Phys. Rev. B} {\bf 80}, 235431 (5 pp).
\bibitem{43} Li G. and Andrei E.Y., 2007. Observation of landau levels of
dirac fermions in graphite. \emph{Nature Phys.} {\bf 3}, 623-627.
\bibitem{44} Miller D.L., Kubista K.D., Rutter G. M., Ruan M., der Heer W.
A., First P. N. and Stroscio A., 2009. Observing the quantization of zero mass carriers in graphene. \emph{Science} {\bf 324}, 924-927.
\bibitem{45} Liang W., Bockrath Marc, Bozovic D., Hafner J.H., Tinkham
M. and Park Hongkun, 2001. Fabryperot interference in a
nanotube electron waveguide. \emph{Nature} {\bf 411}, 665-669.
\bibitem{46} Topsakal M., Cahangirov S., Bekaroglu E. and Ciraci S., 2009.
First-principles study of zinc oxide honeycomb structures. \emph{Phys. Rev. B} {\bf 80}, 235119 (14 pp).
\bibitem{47} Krainara N., Limtrakul J., Illas F. and Bromley S.T., 2011.
Structural and electronic bistability in ZnS single sheets and single-walled nanotubes. \emph {Phys. Rev. B} {\bf 83}, 233305 (4 pp).
\bibitem{48} Dzade N.Y., Obodo K.O., Adjokatse S.K., Ashu A.C.,
Amankwah E., Atiso C.D., Bello A.A., Igumbor E.,
Nzabarinda S.B., Obodo J.T., Ogbuu A.O., Femi O. E.,
Udeigwe J.O. and Waghmare U.V., 2010. Silicene and
transition metal based materials: prediction of a twodimensional
piezomagnet. \emph{J. Phys.: Condens. Mater.} {\bf 22}, 375-502.
\bibitem{49} Michel K.H. and Verberck B., 2009. Theory of elastic and
piezoelectric effects in two-dimensional hexagonal boron
nitride. \emph{Phys. Rev. B} {\bf 80}, 224301 (10 pp).

\end{thebibliography}
\bibliographystyle{elsarticle-num}

\end{document}